\documentclass{ws-p8-50x6-00}
\usepackage{bbbold}
\def\Eins{\mathbbb{1}}%

\def\re{\mathop{\mathrm{Re}}\nolimits}
\def\im{\mathop{\mathrm{Im}}\nolimits}

\newcommand{\bra}[1]{{\langle #1|}} \newcommand{\ket}[1]{{| #1\rangle}}
\newcommand{\bracket}[2]{{\langle #1|#2\rangle}}

\newcommand{\I}{\mathrm{i\,}}
\newcommand{\E}{\mathrm{e\,}}

\newcommand{\ah}{\hat{a}}
\newcommand{\ahd}{\hat{a}^{\dagger}}

\begin{document}

\title{Numerical evaluation of coherent-state path integrals with applications to time-dependent problems}

\author{Bernd Burghardt and Joachim Stolze}

\address{Institut f\"ur Physik, Universit\"at Dortmund, D--44221 Dortmund, Germany, E-mail:~Burghard@cip.physik.uni-dortmund.de}


\maketitle

\abstracts{
We study the application of the coherent-state path integral as a numerical tool for wave-packet propagation. 
The numerical evaluation of path integrals is reduced to a matrix-vector multiplication scheme.
Together with a split-operator technique we apply our method to a time-dependent double-well potential.
}

\section{Introduction and Definitions}

Throughout this paper, we will consider a standard Hamiltonian
\begin{equation}
\hat H=\hat T+\hat V=\frac{{\hat P}^2}{2m}+V(\hat Q)
\label{hamiltonian}
\end{equation}
for a system with one degree of freedom, described by the momentum operator
$\hat P$ and the position operator $\hat Q$.

A coherent state $\ket\alpha$ may be defined by means of harmonic oscillator
creation and annihilation operators $\ahd$ and $\ah$, respectively,
\begin{equation}
\ah := \frac{1}{\sqrt{ 2 \hbar}} (\sqrt{m \omega_{0}}\,\hat Q + 
\I \frac{1}{\sqrt{m \omega_{0}}}\,\hat P)
\label{vernichter},
\end{equation}
through
\begin{equation}
\ket\alpha := \exp(\alpha \ahd - \alpha^* \ah) \ket0\,,\quad \alpha\in\mathbbb{C}
\label{coherent_state},
\end{equation}
where $\ket0 $ is the normalised ground state of the harmonic oscillator $\hat H_{0}=\hbar\omega_{0}\left(\ahd\ah+1/2\right)$, and the exponential
is a displacement operator.
                                %
(Note that the frequency $\omega_{0}$, and hence the characteristic length scale
$(\hbar/m\omega_{0})^{-1/2}$, is completely arbitrary and can be used as an
adjustable parameter.)

The time evolution of a coherent state under $H_{0}$ is simple:
\begin{equation}
  \label{cs_evolution}
  \E^{-\frac{\I t}{\hbar}\hat H_{0}}\ket\alpha =\E^{-\frac{\I \omega_{0}t}{2}} \ket{\alpha\E^{-{\I \omega_{0}t}}}\;.
\end{equation}

For an operator $\hat A$ we define the antinormal symbol $A_{-}(\alpha)$ implicitly by the relation $(d^2 \alpha := d\re\alpha\, d\im\alpha)$
\begin{equation}
  \label{antinormal_symbol}
 \hat A=\int\frac{d^{2}\alpha}{\pi}\ket\alpha\bra\alpha\,A_{-}(\alpha)\;.
\end{equation}

By virtue of the generalised Trotter formula\cite{Che68}
\begin{equation}
  \label{trotter}
  \lim_{N\to\infty}\left(\hat F(t/N)\right)^{N}=\E^{-\I t\,\hat H/\hbar}\;,
\end{equation}
where $\hat F(t)$  is any operator-valued function with the two properties
\begin{eqnarray}
  \label{operator_valued_function}
 \hat  F(t=0)&=&\Eins\nonumber\\
  \dot{\hat F}(t=0)&:=&\lim_{t\to0^{+}}\frac{1}{t}\left(\hat F(t)-\Eins\right)=-\I \hat H/\hbar
\end{eqnarray}
we are able to define the antinormal coherent-state path integral (ACSPI).
However, to exploit the trivial time development of a coherent state under a harmonic oscillator (eq.\ \ref{cs_evolution}), we consider $\hat F$ of a generalised split-operator type:
\begin{equation}
  \label{split_product}
 \hat F(t):= \E^{-\frac{\I t}{2\hbar}\hat H_{0}}\; \hat G(t)\; \E^{-\frac{\I t}{2\hbar}\hat H_{0}}
\end{equation}
with an operator $\hat G$ such that (\ref{operator_valued_function}) holds.
We represent $\hat G$ by eq.\ (\ref{antinormal_symbol}) and define the ACSPI
\begin{eqnarray}
   \label{acspi}
 \lefteqn{\bra{\alpha} \E^{-\I t\,\hat H/\hbar}\ket{\alpha^{\prime}} :=\E^{-\I \omega_{0}t/2}\cdot}\\ 
& & \lim_{N\to\infty} \int \frac{d^{2}\alpha_{1}}{\pi}\cdots\frac{d^{2}\alpha_{N}}{\pi} \prod_{\nu=0}^{N}\bracket{\alpha_{\nu}\E^{+\I\omega_{0} t/2}}{\alpha_{\nu+1}\E^{-\I\omega_{0} t/2}}
\prod_{\nu=1}^{N}G_{-}(\alpha_{\nu};t/N)\nonumber
\end{eqnarray}
$\alpha\equiv\alpha_{0}\E^{+\I\omega_{0} t/2}, \alpha^{\prime}\equiv\alpha_{N+1}\E^{-\I\omega_{0} t/2}$.

Here we use 
\begin{equation}
  \label{split_symbol}
 \hat G(t):= \sum_{n=0}^{K}\frac{(-\I t)^{n}}{n!}(\hat H_{1}/\hbar)^{n}
\end{equation}
where $\hat H_{1}=V(\hat Q)-\frac{m\omega_{0}^{2}}{2}{\hat Q}^{2}$ is the anharmonic part of the Hamiltonian $\hat H$ and $K\in\left\{4,5,\dots,10\right\}$.

An analogous normal coherent-state path integral may also be defined,\cite{BES98} but will not be considered here.

\section{Numerical evaluation}
\label{sec:numerics}

For numerical evaluation of the ACSPI (eq.\ \ref{acspi}) we have to stop at a finite (Trotter-)number $N$ of integrations and we perform each integration by a quadrature formula 
\begin{equation}
  \label{quadrature}
  \int\frac{d^2\alpha}{\pi}f(\alpha)\approx \sum_{j}w_{j}\, f(\alpha_{j})
\end{equation}
with fixed sets of abscissas $\alpha_{j}$ and weights $w_{j}>0$, and defining
a matrix $P(t/N)$ and a vector $v$ by their elements
\begin{eqnarray}
  \label{anti_prop_matrix}
  P_{ij}&=&\sqrt{w_{i}w_{j}}\,\E^{-\I\frac{\omega t}{2N}}\,\bracket{\alpha_{i} \E^{+\I\frac{\omega t}{N}}}{\alpha_{j}}\; G_{-}(\alpha_{i} \E^{+\I\frac{\omega t}{2N}}; t/N),
\end{eqnarray}
and
\begin{equation}
  \label{state_vector}
  v_{j}:=\sqrt{w_{j}}\bracket{\alpha_{j}}{\psi},
\end{equation}
the discretized version of a single time step propagation becomes a
matrix-vector-multiplication
\begin{equation}
  \label{discrete_prop}
  v_{i}^{\prime}:=\sqrt{w_{i}}\bra{\alpha_{i}}\E^{-\frac{\I t}{\hbar N}\hat H}\ket\psi\simeq \sum_{j}P_{ij}(t/N)\,v_{j}.
\end{equation}
The vector $v^{\prime}$ represents the wave packet propagated by the time $t/N$.

For time-dependent potentials we choose the time step small enough to treat the potential as time-independent during the time $t/N$.

\begin{figure}[tbp]
  \begin{center}
    \leavevmode
    \centerline{ \epsfxsize=11cm \epsfbox{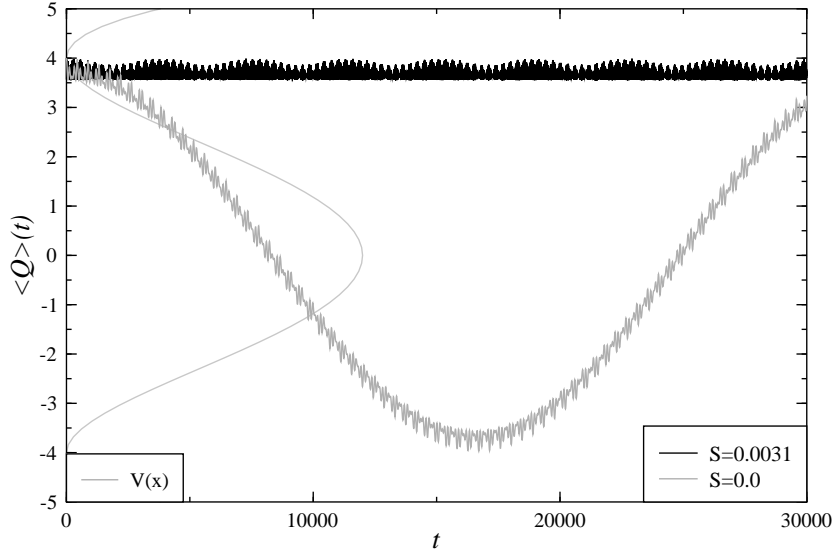} }
    \caption{The $Q$-expectation value as a function of time for a particle in the potential of equation (\ref{potential}) is shown.
In absence of an external field $(S=0)$ the particle is tunneling through the barrier due to nearly degenerate energy eigenvalues.
The destruction of tunneling for $S\not=0$ is due to degenerate quasi-energies in the Floquet picture.\protect\cite{GDJ*91}
}  \label{figure}
  \end{center}
\end{figure}

\section{Application to tunneling phenomena}
\label{sec:application}
We apply our method to a symmetric double well potential with an external time-periodic linear potential:
\begin{equation}
  \label{potential}
  V(\hat Q)=\frac{m\omega_{0}^{2}}{8 Q_{0}^{2}}\,\left({\hat Q}^{2}-Q_{0}^{2}\right)^{2}+S \sin\left(\omega t\right)\,\hat Q
\end{equation}
The time-independent case ($S=0$) shows the phenomena of tunneling (see fig.\ \ref{figure}): A wave packet starting in one well moves through the barrier into the other well.
However, Gro{\ss}mann et al.\cite{GDJ*91} observed that application of a time-periodic linear potential with the right strength $S$ and frequency $\omega$ can localize the particle in one well.

In figure \ref{figure} we show the $x$-expectation value of the wave packet as a function of time.
Without external field the particle needs about 2500 elementary (single well) oscillation periods to tunnel from one well to the other.
However, application of a field with strength $S=0.0031$ and frequency $\omega=0.01$ suppresses the tunneling process. 
It is remarkable that this problem has three different time scales: The vibrational period $T=2\pi/\omega_{0}$ around one minimum, the period $T=200\pi/\omega_{0}$ of the driving force, and the tunneling period $T\approx2\cdot10^{5}/\omega_{0}$.
This shows the capability of our method to give reliable results for problems with time scales ranging over several orders of magnitude.

\section*{Acknowledgements}
This work was supported by the Deutsche Forschungs\-gemein\-schaft through the
Schwer\-punkt\-pro\-gramm: Zeit\-abh\"angige Ph\"a\-no\-me\-ne und Methoden in
Quantensystemen der Physik und Chemie.


\newcommand{\newblock}{}
\bibliography{bbas_def,bbase}



\end{document}